\documentclass[11pt]{article}
\usepackage{blois2002,epsfig}
\begin{document}
\vspace*{4cm}
\title{DETERMINING STRONG AND WEAK PHASES IN
$B_d \rightarrow D^{(*)\pm}D^{(*)\mp}$ DECAYS}

\author{Z.Z. XING}

\address{Institute of High Energy Physics, P.O. Box 918 (4),
Beijing 100039, China}

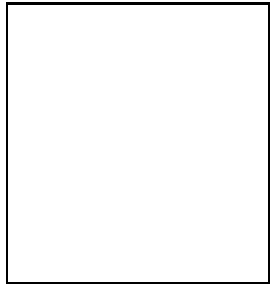
\maketitle\abstracts{
We show that both the weak phase $\beta$ and the strong phase $\delta_d$
can be determined from the time-dependent measurement of 
$B_d \rightarrow D^{*\pm}D^{\mp}$ decays, whose final states are 
non-CP eigenstates. It is also possible to extract $\beta$ from
$B_d \rightarrow D^{*\pm}D^{*\mp}$ transitions without doing the
angular analysis. Possible final-state rescattering effects in
$B_d \rightarrow D^{(*)\pm}D^{(*)\mp}$ channels are discussed by means of
the isospin analysis. We emphasize that it is worthwhile to check
whether the naive factorization approximation works or not for
such $B$-meson decay modes into two heavy charmed mesons.}

\section{Introduction}

Weak decay modes $B_d \rightarrow D^+D^-$, $D^{*+}D^-$, 
$D^+D^{*-}$ and $D^{*+}D^{*-}$ are interesting 
for the study of CP violation and final-state interactions
at $B$-meson factories. The experimental result for the
branching fraction of $B_d \rightarrow D^{*+}D^{*-}$ is
${\cal B} (D^{*+}D^{*-}) = (9.9^{+4.4}_{-3.5}) \times 10^{-4}$ \cite{PDG}.
Recently the Belle Collaboration has reported the first
measurement of $B_d\rightarrow D^{\pm}D^{*\mp}$ decays. 
The sum of their branching fractions is \cite{Belle}
${\cal B}(D^+D^{*-} \oplus D^-D^{*+}) = 
(1.17 \pm 0.26^{+0.22}_{-0.25}) \times 10^{-3}$
(full reconstruction method) or
$(1.48 \pm 0.38^{+0.28}_{-0.31}) \times 10^{-3}$ 
(partial reconstruction method).
We hope that a measurement of $B_d \rightarrow D^+D^-$ will
soon be available.

The $B_d \rightarrow D^{(*)+}D^{(*)-}$ channels are associated 
with the weak CP-violating phase
\begin{equation}
\beta \; \equiv \; \arg \left (- \frac{V^*_{tb}V_{td}}
{V^*_{cb}V_{cd}} \right ) \; ,
\end{equation}
where $V_{ij}$ (for $i=u,c,t$ and $j=d,s,b$) are elements of
the Cabibbo-Kobayashi-Maskawa matrix. A determination of
$\beta$ from CP-violating asymmetries of $B_d \rightarrow
D^{(*)+}D^{(*)-}$ transitions will be useful, not only to cross-check the
extraction of $\beta$ from $B_d \rightarrow J/\psi K_S$, but
also to shed some light on the relevant penguin effects and final-state
interactions. In addition, it is important to test whether the
naive factorization approximation works or not for such $B$ decay modes 
into two heavy charmed mesons.

\section{Determining strong and weak phases in
$B_d\rightarrow D^{\pm}D^{*\mp}$ decays}

The transitions $B^0_d \rightarrow D^{*\pm}D^{\mp}$ can occur through
both tree-level and loop-induced (penguin) quark diagrams,
The penguin contribution to the overall amplitude of each decay mode
is negligible \cite{Xing98}. In this good approximation, one may define 
two interference quantities between decay amplitudes and 
$B^0_d$-$\bar{B}^0_d$ mixing:
\begin{eqnarray}
\lambda_{D^{*+}D^-} & \equiv & \frac{q^{~}_d}{p^{~}_d} 
\cdot \frac{A(\bar{B}^0_d\rightarrow D^{*+}D^-)}
{A(B^0_d\rightarrow D^{*+}D^-)} \; =\;
\frac{V^*_{tb}V_{td}}{V_{tb}V^*_{td}}\cdot
\frac{V_{cb}V_{cd}^*}{V^*_{cb}V_{cd}} 
~ \zeta_d ~ e^{{\rm i}\delta_d} \; =\;
\zeta_d ~ e^{{\rm i} (\delta_d + 2\beta)} \; , \nonumber \\
\bar{\lambda}_{D^{*-}D^+} & \equiv & \frac{p^{~}_d}{q^{~}_d} 
\cdot \frac{A(B^0_d\rightarrow D^{*-}D^+)}
{A(\bar{B}^0_d\rightarrow D^{*-}D^+)} \; =\;
\frac{V_{tb}V^*_{td}}{V^*_{tb}V_{td}}\cdot
\frac{V^*_{cb}V_{cd}}{V_{cb}V^*_{cd}} 
~ \zeta_d ~ e^{{\rm i}\delta_d} \; =\;
\zeta_d ~ e^{{\rm i} (\delta_d - 2\beta)} \; ,
\end{eqnarray}
where $q^{~}_d/p^{~}_d = (V^*_{tb}V_{td})/(V_{tb}V^*_{td})$
describes the $B^0_d$-$\bar{B}^0_d$ mixing
phase in the box-diagram approximation, $\zeta_d$ and
$\delta_d$ denote the ratio of the real hadronic matrix elements and
the strong phase difference between 
$\bar{B}^0_d\rightarrow D^{*+}D^-$ and $B^0_d\rightarrow D^{*+}D^-$.
In the naive factorization approximation, we have
$\zeta_d = [f_D \cdot A^{B_dD^*}_0(m^2_D)]/[f_{D^*} \cdot
F^{B_dD}_1 (m^2_{D^*})] \approx 1.04$,
where the relevant decay constants and formfactors are
self-explanatory. Comparing the experimental and theoretical results of
$\zeta_d$ will provide a clean test of the factorization
hypothesis for neutral-$B$ decays into two heavy charmed mesons.

The imaginary parts of $\lambda_{D^{*+}D^-}$ and
$\bar{\lambda}_{D^{*-}D^+}$ are of particular interest for
the study of CP violation. It should be noted, however, that 
${\rm Im}\lambda_f$ and ${\rm Im} \bar{\lambda}_{\bar{f}}$ (for
$f = D^{*+}D^-$) themselves are not CP-violating observables! Only
their difference ${\rm Im} (\lambda_f - \bar{\lambda}_{\bar{f}})$, 
which will vanish for $\beta =0$ or $\pi$, 
measures the CP asymmetry.
The time-dependent rates of $B_d\rightarrow D^\pm D^{*\mp}$ modes
read as \cite{Xing98}
\begin{eqnarray}
{\cal R} \left [ \stackrel{\langle - \rangle}{B^0_d}(t)
\rightarrow D^{*+}D^- \right ] 
& \propto & \left [\frac{1 +\zeta^2_d}{2} \stackrel{\langle - \rangle}
{+} \frac{1 -\zeta^2_d}{2} \cos (x_d \Gamma_d t) 
\stackrel{\langle + \rangle}{-} 
\zeta_d \sin (\delta_d + 2\beta) \sin (x_d \Gamma_d t)
\right ] , \nonumber \\
{\cal R} \left [ \stackrel{\langle - \rangle}{B^0_d}(t)
\rightarrow D^{*-}D^+ \right ] 
& \propto & \left [ \frac{1 +\zeta^2_d}{2} \stackrel{\langle + \rangle}
{-} \frac{1 -\zeta^2_d}{2} \cos (x_d \Gamma_d t) 
\stackrel{\langle - \rangle}{+} 
\zeta_d \sin (\delta_d - 2\beta) \sin (x_d \Gamma_d t)
\right ] ;
\end{eqnarray}
where $x_d \approx 0.7$ is the $B^0_d$-$\bar{B}^0_d$ mixing parameter,
and $\Gamma_d$ denotes the $B_d$ decay width. Then
we may extract the weak phase $\beta$ and the
strong phase $\delta_d$ up to a four-fold ambiguity:
\begin{eqnarray}
\sin^2 (2\beta) & = & \frac{1}{2} \left [(1 -S_+S_-)
~ \pm ~ \sqrt{(1-S^2_+) (1-S^2_-)} \right ] \; , 
\nonumber \\
\sin^2 \delta_d & = & \frac{1}{2} \left [(1 +S_+S_-)
~ \pm ~ \sqrt{(1-S^2_+) (1-S^2_-)} \right ] \; , 
\end{eqnarray}
where $S_{\pm} \equiv \sin (\delta_d \pm 2\beta)$.
Indeed only a two-fold ambiguity in $\sin(2\beta)$
exists, as $\sin (2\beta) >0$ has been experimentally verified
within the standard model \cite{PDG}.
If final-state interactions were insignificant in the decay modes
under discussion, $\delta_d$ might not deviate too much from zero.
In this case, $S_+ \approx - S_-$ would be a good approximation. 

\section{Extracting $\beta$ from $B_d\rightarrow D^{*\pm}D^{*\mp}$ decays
without angular analysis}

A comparison between the value of $\sin 2\beta$ 
to be determined from $B_d\rightarrow D^{*+}D^{*-}$ and that
already measured in $B_d\rightarrow J/\psi K_{\rm S}$ is no doubt 
important, as it may cross-check the consistency of the standard-model 
predictions. Towards this goal, a
special attention has to be paid to possible uncertainties
associated with the CP asymmetry in $B_d\rightarrow D^{*+}D^{*-}$. 
One kind of uncertainty comes from the penguin contamination,
as the weak phase of the penguin amplitude is 
quite different from that of the tree-level amplitude.
Another kind of uncertainty arises from the $P$-wave dilution,
because the final state $D^{*+}D^{*-}$ is composed
of both the CP-even ($S$- and $D$-wave) 
and the CP-odd ($P$-wave) configurations. Of course an
analysis of the angular distributions of $B^0_d$ vs
$\bar{B}^0_d\rightarrow D^{*+}D^{*-}$ transitions allows us
to distinguish between the CP-even and CP-odd 
contributions \cite{AA}. Here we like to emphasize that
the direct measurement of  $\beta$ can be made in 
$B_d\rightarrow D^{*+}D^{*-}$ decays without doing the angular 
analysis \cite{Pham}.  

Taking the $P$-wave dilution and the penguin contamination
into account, one may write
the characteristic measurable of indirect CP
violation in $B_d\rightarrow D^{*+}D^{*-}$ as follows
\begin{equation}
\Delta_d \equiv {\rm Im} \left (
\frac{V^*_{tb}V_{td}}{V_{tb}V^*_{td}} \cdot
\frac{\langle D^{*+}D^{*-}|{\cal H}_{\rm eff}|
\bar{B}^0_d\rangle}
{\langle D^{*+}D^{*-}|{\cal H}_{\rm eff}|B^0_d\rangle} \right ) 
= P_d \left ( 1 - Q_d \right )  \sin 2\beta \; ,
\end{equation}
where $P_d$ and $Q_d$ 
represent the $P$-wave dilution factor and the penguin-induced
correction, respectively. 
With the help of the effective weak Hamiltonian, the naive factorization
approximation and the heavy quark symmetry, we obtain \cite{Pham}
\begin{eqnarray}
P_d & = & \frac{m^3_B - 3 m^{~}_B m^2_{D^*} + 10 m^3_{D^*}}
{m^3_B + m^{~}_B m^2_{D^*} + 2 m^3_{D^*}} \;\; ,
\nonumber \\
Q_d & = & \frac{c_y + c_z}{c_x} \cdot \frac{\cos 2\beta}{\cos\beta}
\cdot \left | \frac{V_{tb}V_{td}}{V_{cb}V_{cd}} \right | 
\;\; ,
\end{eqnarray}
where $c_x \approx 1.045$, $c_y \approx -0.031$ and
$c_z \approx -0.0014$ are the effective Wilson coefficients.
Typically taking $\beta = 26^\circ$, which is favored by current
BaBar and Belle data \cite{Beta}, we find
$P_d \approx 0.89$ and $Q_d \approx -0.021$.
This result indicates that the penguin contamination in $\Delta_d$
is negligibly small, while the $P$-wave dilution to $\Delta_d$
should be taken seriously. 

It is worth remarking that the approach advocated here may be
complementary to the angular analysis considered in the literature.
Hopefully both will soon be confronted with the new data from $B$-meson
factories.
 
\section{Final-state rescattering effects in 
$B_d\rightarrow D^{(*)\pm}D^{(*)\mp}$ decays}

The effective weak Hamiltonian responsible for 
$B^-_u\rightarrow D^-D^0$, $\bar{B}^0_d\rightarrow D^+D^-$
and $\bar{B}^0_d\rightarrow D^0\bar{D}^0$ decay modes
has the isospin structure $|1/2, -1/2\rangle$.
The decay amplitudes of these transitions can be written in terms 
of the isospin amplitudes \cite{Sanda}:
\begin{eqnarray}
A^{+-} & \equiv & \langle D^+D^- |{\cal H}_{\rm eff}| B^0_d\rangle \; = \;
\frac{1}{2} \left ( A_1 ~ + ~ A_0 \right ) \; , \nonumber \\
A^{00} & \equiv & \langle D^0\bar{D}^0 |{\cal H}_{\rm eff}| B^0_d\rangle 
\; = \;
\frac{1}{2} \left ( A_1  ~ - ~ A_0  \right ) \; , \nonumber \\
A^{+0} & \equiv & \langle D^+\bar{D}^0 |{\cal H}_{\rm eff}| B^+_u\rangle 
\; = \; A_1 \; ,
\end{eqnarray}
where $A_1$ and $A_0$ are the isospin
amplitudes with $I=1$ and $I=0$, respectively. 
Clearly the isospin relation $A^{+-} + A^{00} = A^{+0}$ holds,
and it corresponds to a triangle in the complex plane.
Denoting $A_0/A_1 \equiv z e^{{\rm i}\theta}$, we obtain
\begin{eqnarray}
z = \sqrt{\frac{2 \displaystyle \left ( |A^{+-}|^2 + |A^{00}|^2 \right )}
{|A^{+0}|^2} - 1} \; , ~~~~
\theta = \arccos \left ( \frac{|A^{+-}|^2 - |A^{00}|^2}
{z |A^{+0}|^2} \right ) \; ; 
\end{eqnarray}
If $z=1$ and $\theta =0$, for example, we find that $|A^{00}|=0$, i.e.,
the decay mode $B^0\rightarrow D^0\bar{D}^0$ is forbidden. 
One may get similar isospin relations for the decay modes 
$B^+_u\rightarrow D^+\bar{D}^0$, $B^0_d\rightarrow D^+D^-$ and
$B^0_d\rightarrow D^0\bar{D}^0$.

It is worth mentioning that the same isospin analysis can be done for 
$B\rightarrow D\bar{D}^*$ and $B\rightarrow D^*\bar{D}$ decays. Of course,
the isospin parameters $z$ ($\bar{z}$) and $\theta$ ($\bar{\theta}$) in
$B\rightarrow D\bar{D}$, $D\bar{D}^*$ and $D^*\bar{D}$ may be different 
from one another due to their different final-state interactions. 
As for $B\rightarrow D^*\bar{D}^*$,
the same isospin relations hold separately for the decay amplitudes
with helicity $\lambda=-1$, $0$, or $+1$ \cite{Xing00}.

The time-independent measurements of those decay modes
mentioned above allow us to construct the relevant isospin triangles. 
Consequently the isospin parameters $z$ ($\bar{z}$)
and $\theta$ ($\bar{\theta}$) are extractable in the absence of any 
time-dependent measurement. If the branching ratios of $B^0_d\rightarrow 
D^0\bar{D}^0$ and $\bar{B}^0_d\rightarrow D^0\bar{D}^0$ are too small to be
observable, then large cancellation between the isospin amplitudes $A_1$
($\bar{A}_1$) and $A_0$ ($\bar{A}_0$) must take place. In the case that
$B^0_d\rightarrow D^+D^-$ and $B^+_u\rightarrow D^+\bar{D}^0$ have been
measured earlier than $B^0_d\rightarrow D^0\bar{D}^0$, a lower bound on the
rate of the latter decay mode is model-independently achievable from 
the isospin relations obtained in Eq. (7). Since $\cos\theta \leq 1$, 
we get from Eqs. (7) and (8) that
\begin{equation}
{\cal B} (B^0_d\rightarrow D^0\bar{D}^0) \geq \left [ \sqrt{\frac{
{\cal B} (B^0_d\rightarrow D^+D^-)}{{\cal B} 
(B^+_u\rightarrow D^+\bar{D}^0)}} - 1
\right ]^2 {\cal B} (B^+_u\rightarrow D^+\bar{D}^0) \; ,
\end{equation}
where tiny isospin-violating effects induced by the mass difference 
$m^{~}_{D^0} - m^{~}_{D^-}$ and the life time difference $\tau^{~}_{B_d}
- \tau^{~}_{B_u}$ have been neglected. This bound is useful to 
set a limit for the results of 
${\cal B} (B^0_d\rightarrow D^0\bar{D}^0)$ obtained
from specific models of hadronic matrix elements.
Similarly one can find the lower bounds for the 
branching ratios of $B^0_d\rightarrow D^{*0}\bar{D}^0$, $D^0\bar{D}^{*0}$ 
and $D^{*0}\bar{D}^{*0}$.

\section{Concluding remarks}

Some conclusions can be drawn from our results:
(a) $B_d\rightarrow D^{\pm}D^{*\mp}$ modes are useful to
determine the weak CP-violating phase $\beta$ and the strong phase
shift $\delta_d$; (b) $\beta$ can also be determined from 
$B_d\rightarrow D^{*\pm}D^{*\mp}$ transitions without doing the angular
analysis; (c) it is worthwhile to investigate final-state rescattering 
effects in $B\rightarrow D^{(*)}\bar{D}^{(*)}$ channels, and to check
whether the naive factorization approximation works or not for such
$B$ decays into two heavy charmed mesons.

Similar analyses can be done for $B_s\rightarrow D^{(*)\pm}_sD^{(*)\mp}_s$ 
decay modes.

\section*{Acknowledgments}

I would like to thank J. Tran Thanh Van for his encouragement and
hospitality. I am also grateful to L. Oliver for
useful communications and R. Aleksan for helpful discussions.

\end{document}